\documentclass[aps,prl,twocolumn,superscriptaddress]{revtex4-1}

\usepackage{mathtools}
\usepackage{amsmath,bm}
\usepackage{amssymb}
\usepackage{subfig}
\usepackage{booktabs}			
\usepackage{caption}
\usepackage{tabularx}
\usepackage{hyperref}
\usepackage{graphicx}
\usepackage{setspace}
\usepackage{color}

\begin{document}


\title{Experimental observation of non-reciprocal band-gaps in a space-time modulated beam using a shunted piezoelectric array}

\author{J. Marconi}
\affiliation{Department of Mechanical Engineering, Politecnico di Milano, 20156, Italy}
\author{E. Riva}
\email[]{emanuele.riva@polimi.it}
\affiliation{Department of Mechanical Engineering, Politecnico di Milano, 20156, Italy}
\author{M. Di Ronco}
\affiliation{Department of Mechanical Engineering, Politecnico di Milano, 20156, Italy}
\author{G. Cazzulani}
\affiliation{Department of Mechanical Engineering, Politecnico di Milano, 20156, Italy}
\author{F. Braghin}
\affiliation{Department of Mechanical Engineering, Politecnico di Milano, 20156, Italy}
\author{M. Ruzzene}
\affiliation{Department of Mechanical Engineering, University of Colorado, Boulder, CO 80309, USA}


\date{\today}

\begin{abstract}
In this work we experimentally achieve 1 kHz-wide directional band-gaps for elastic waves spanning a frequency range from approximately 8 to 11 kHz. One-way propagation is induced by way of a periodic waveguide consisting in an aluminum beam partially covered by a tightly packed array of piezoelectric patches. The latter are connected to shunt circuits and switches which allow for a periodic modulation in time of the cell properties. A traveling stiffness profile is obtained by opportunely phasing the temporal modulation of each active element, mimicking the propagation of a plane wave along the material, therefore establishing unidirectional wave propagation at bandgap frequencies. 
\end{abstract}


\maketitle

\begin{figure*}[t!] 
\subfloat[]{\includegraphics[width=.55\textwidth]{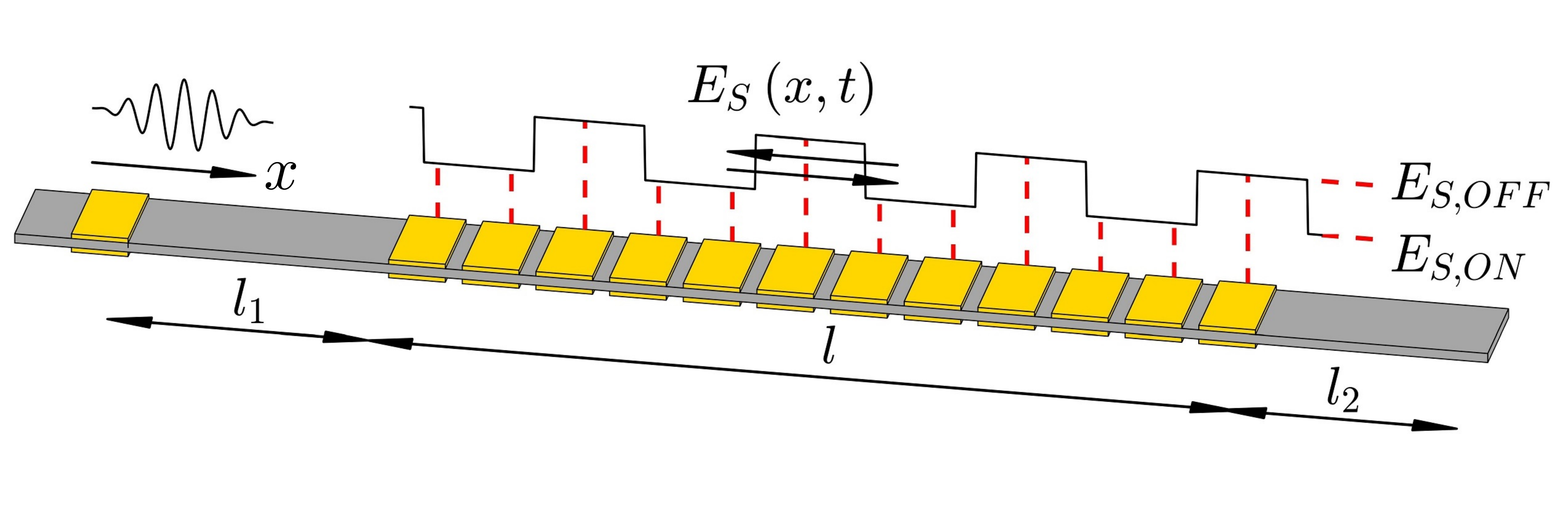} \label{fig:1a}} 
\subfloat[]{\includegraphics[width=.45\textwidth]{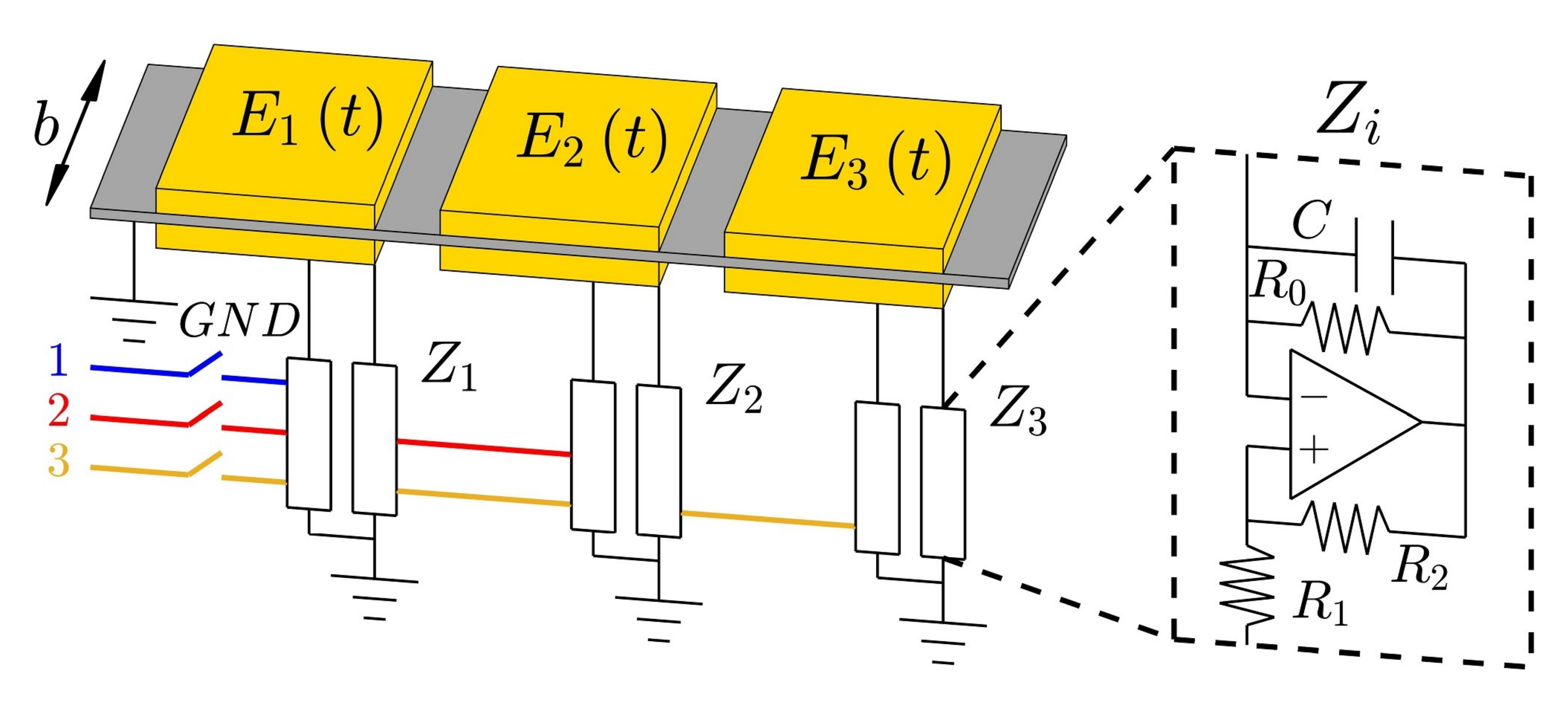} \label{fig:1b}} 
\caption{Schematic of the electro-mechanical beam, showing (a) the excitation patch providing the tone burst, the array of patches of the active domain and (b) a close-in of the ST cell including the shunted series NC. High (low) stiffness is obtained by opening (closing) the switches connecting the power supply to the operational amplifiers.}
\label{fig:1}
\end{figure*}

Nonreciprocal devices have been pursued in various research domains and physical platforms, including quantum \cite{hasan2010colloquium}, electromagnetic \cite{khanikaev2013photonic,chamanara2017optical}, acoustic \cite{cummer2016controlling,li2019nonreciprocal,karkar2019broadband} and elastic \cite{vila2017bloch,nassar2017modulated,wallen2019nonreciprocal} media. These devices support wave propagation from a point (A) to one other (B), but not vice-versa, opening up new possibilities for the control of energy flow with unprecedented performance in communication systems \cite{huber2016topological}, unidirectional insulators \cite{trainiti2016non} and converters \cite{yi2018reflection,oudich2019space}, among others. Important contributions in the context of one-way phonon transport have been formulated by Fleury et. al. \cite{fleury2014sound,fleury2015subwavelength}, demonstrating directional wave manipulation in acoustic cyrculator devices. Also, elastic and acoustic directional waveguides have been conceived and physically realized, in analogy with the \textit{Quantum Hall effect} (QHE), achieving back-scattering immune and one-way topological edge states \cite{khanikaev2015topologically,chen2019mechanical,wang2015topological,ni2015topologically,nash2015topological}. 
Other approaches to nonreciprocity leverage nonlinear phenomena \cite{Mojahed2019,Fronk2019}, metastability \cite{wu2018metastable}, bifurcation and chaos \cite{boechler2011bifurcation} which are particularly attractive solutions due to the presence of solely passive elements. However, the exploitation of nonlinear dynamics usually requires high wave amplitudes, thus making the physical realization impractical for compact devices.\\
An effective platform to break reciprocity is offered by space-time modulated systems \cite{nassar2017non,riva2019generalized}. 
Notable recent examples have employed programmable magnetic lattice elements \cite{chen2019nonreciprocal} and magnetic springs \cite{wang2018observation}.\\
In this work we experimentally investigate non-reciprocity in a phononic beam, where spatial and temporal modulations are induced upon electric control of equivalent elastic properties. Namely, the spatial modulation is induced by bonding a pattern of piezoelectric elements on a passive substrate, which effectively alter the Young-s modulus of the waveguid through negative capacitance shunts \cite{trainiti2019time}, which are manipulated in time through a switching logic. This enables the formation of a traveling stiffness profile, which produces an asymmetric dispersion relation, which is a hallmark of non-reciprocity.\\
As shown in \cite{trainiti2019time}, the proposed configuration an effective mean to test non-reciprocity of spatio-temporally modulated media, and may also be adopted as a flexible platform to explore other phenomena associated with temporal and spatio-temporal modulation, among which parametric amplification \cite{trainiti2019time}, conversion \cite{yi2018reflection}, and topological edge-to-edge pumping \cite{kraus2012topological,grinberg2019robust}.\\
\indent
We consider the electro-mechanical beam illustrated in Fig. \ref{fig:1}, which is made of an aluminum substrate having cross section $b\times H=20\;{\rm mm}\times1\;{\rm mm}$ and total length $L=2400\;{\rm mm}$. 
\begin{figure*}[t!] 
	\subfloat[]{\includegraphics[width=0.43\textwidth]{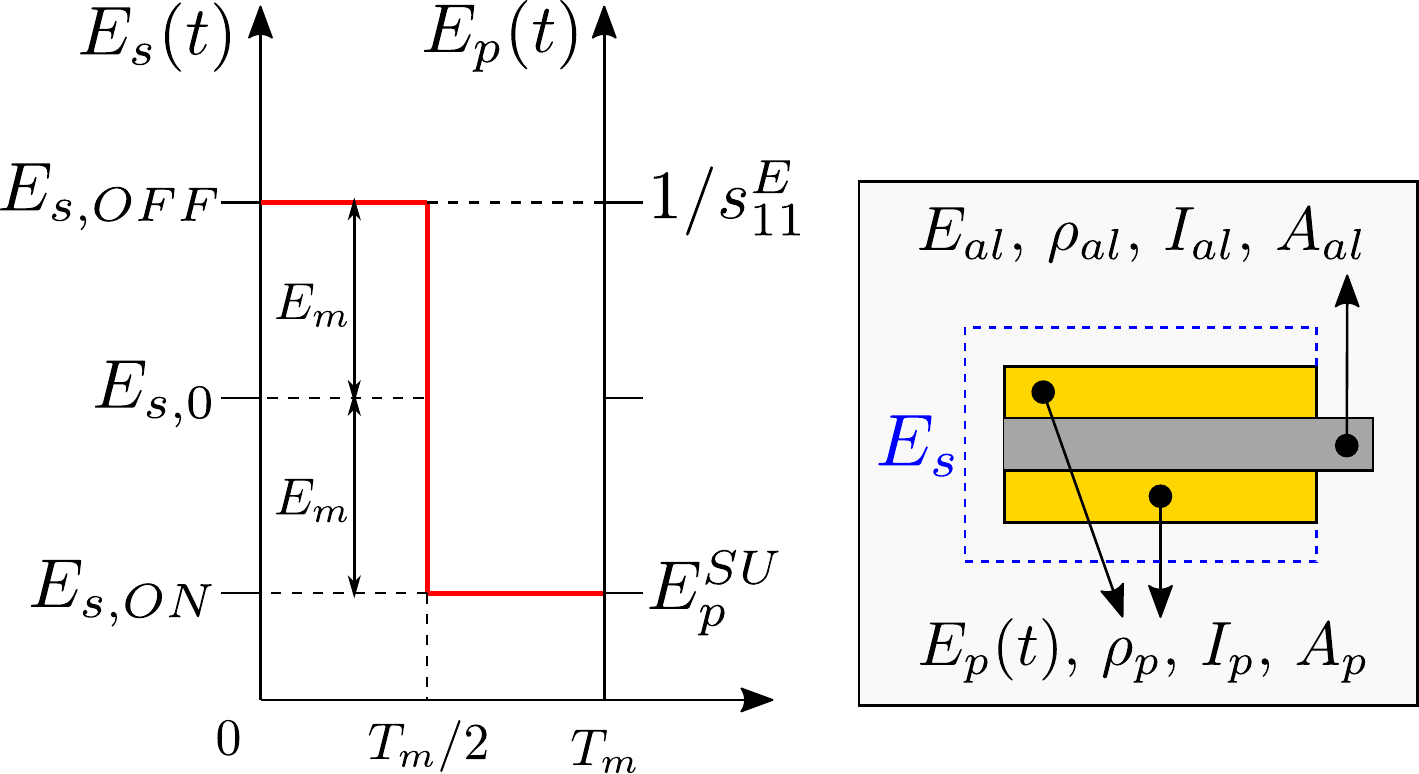} \label{fig:2a}}
	\subfloat[]{\includegraphics[width=0.57\textwidth]{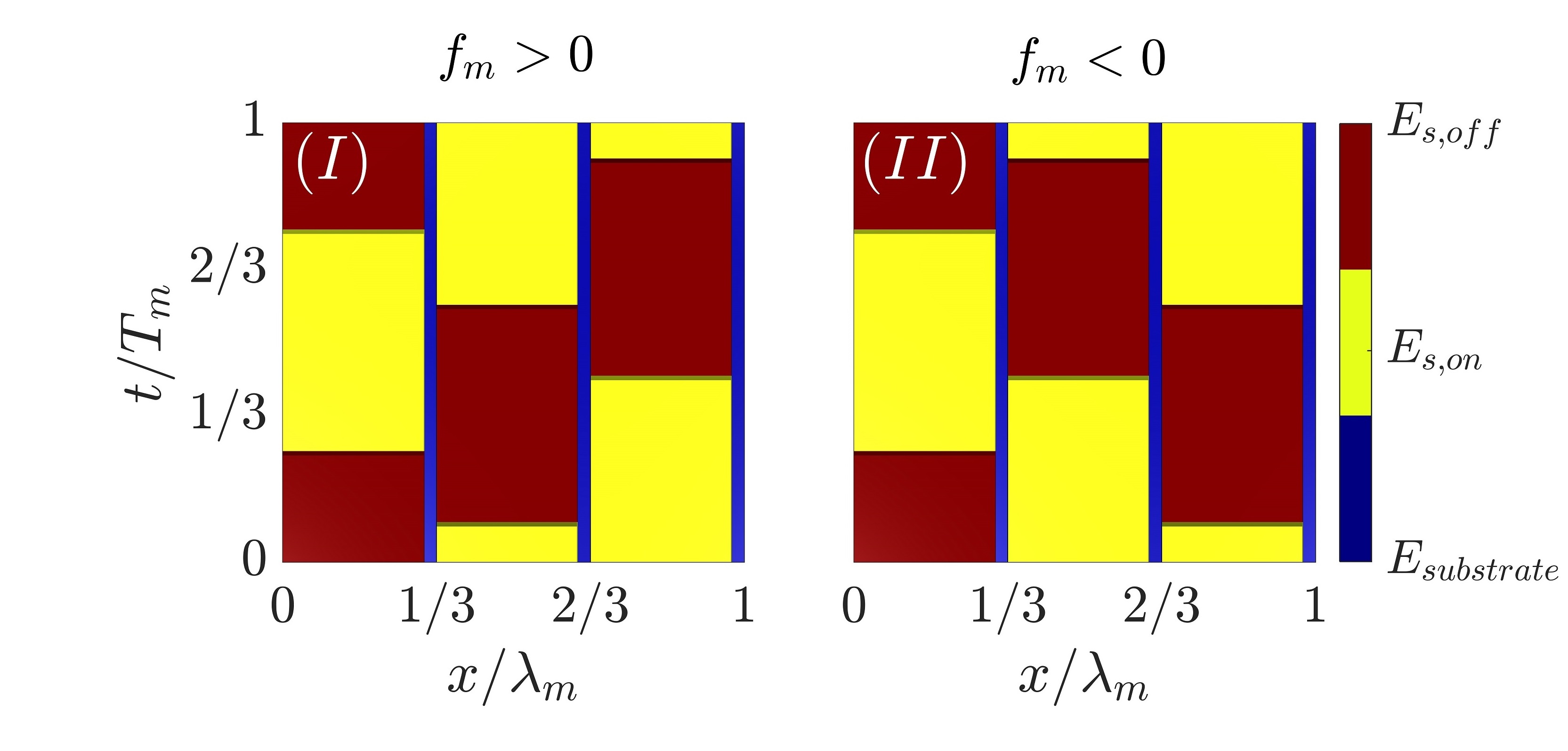} \label{fig:2b}}	
	\label{Fig:2}
	\caption{(a) Synoptic scheme and notation of stiffness for the patch ($E_p$), the sandwich ($E_s$) and of relevant modulation parameters. (b) Forward (${\MakeUppercase{\romannumeral 1 }}$) and backward (${\MakeUppercase{\romannumeral 2 }}$) Young's modulus profiles in the ST-cell domain $D$.}
\end{figure*}
An array of piezoelectric patches, separated by $2\;{\rm mm}$ from each other, is placed approximately at the beam's mid-span shifted by $l_1=690\;{\rm mm}$ and $l_2=1134\;{\rm mm}$ from the left and right boundaries in order to prevent reflections and waves interference during the transient analysis. For the same reason, the system is equipped with absorbing boundaries, obtained by covering the two clamped ends with mastic tape. The piezoelectric active domain is characterized by a length $l=576\;{\rm mm}$ and it consists of $24$ pairs of patches ($\rho_p=7.9\;{\rm kg/dm^3},\;E_p=62\;{\rm GPa}$) of size $b\times l_p\times h_p=20\times22\times1\;{\rm mm}$, bonded on opposite surfaces. Each patch is connected to a shunt circuit emulating a series negative capacitance (NC), for a total of $48$ shunts, which provide an effective stiffness reduction to the beam section when the circuit is closed \cite{trainiti2019time}. Finally, a pair of patches bonded to the left end close to the clamp provide excitation. \\
A group of three patches pair defines a spatio-temporal (ST) cell, arranged in a configuration represented in Fig. \ref{fig:1b}. The effective sandwich stiffness of each ST cell is modulated following the law $E_{s,k}\left(t\right)=E_{s,0}\left[1+\alpha_m\text{sign}\left(\cos\left({2\pi f_mt+\left(k-1\right)\pi/3}\right)\right)\right]$, with $k=1,2,3$ denoting the sub-cell number, while $\pi/3$ is the phase shift between the three consecutive active elements. Also, $\alpha_m$ defines the amplitude of modulation, and $f_m=1/T_m$ is the modulation frequency, where $T_m$ is the modulation period. \\
In the case at hand, the stiffness law is practically determined by the electrical boundary conditions of the piezoelectric patches. Periodically switching the NC circuits OFF and ON alternates the shunt impedance $Z^{SU}_{OFF}=0$ and $Z^{SU}_{ON}=-1/(i\omega C_N)$, being $C_N=CR_2/R_1$ the equivalent value of the synthetic NC shunt circuit. In turn, this impedance change alternates the effective Young's moduli $E_{p}=1/s^E_{11}$ (with $s^E_{11}$ short-circuit mechanical compliance of the patch in 31-operation mode) and $E_{p}^{SU}=E_p\left(C_N-C_p^T\right)/\left(C_N-C_p^S\right)$ when the shunt is turned OFF and ON, respectively. $C_p^{T(S)}$ is the piezo capacitance under stress ($T$) and strain ($S$) free conditions. NC circuit data is collected in Tab. \ref{tab:01}. 
\begin{table}[b]
	\begingroup\makeatletter\def\f@size{9}\check@mathfonts
	\centering
	\begin{tabular}{llll}
		\hline
		\textbf{Name}&\textbf{Value}&\textbf{Units}&\textbf{Description}\\
		$R_1$         & $7.5$            & ${\rm k}\Omega$ &\textit{$-$}\\ \hline
		$R_2$         & $13.7$           & ${\rm k}\Omega$&\textit{$-$}\\ \hline
		$R_0$         & $1000$          & ${\rm k}\Omega$&\textit{bias resistance}\\ \hline
		$C$           & $4.4$            & ${\rm nF}$& \textit{NC capacitance}\\ \hline
		$C_p$         & $6.7\div7$     & ${\rm nF}$& \textit{piezo patch capacity} \\ \hline
		$d_{31}$      & $-1740$          & ${\rm pm/V}$& \textit{piezo strain coefficient} \\ \hline
		$k_{31}$      & $0.351$          & $-$& \textit{piezo coupling coefficient}\\\hline
	\end{tabular}
	\caption{NC circuit parameters.}
	\label{tab:01}
	\endgroup
\end{table}
We now define $E_{s,ON/OFF}$ as biased and unbiased sandwich stiffnesses, which account for both piezo and substrate, as shown by the schematic in Fig. \ref{fig:2a}. 
\begin{figure*}[t!]
	\includegraphics[width=1.0\textwidth]{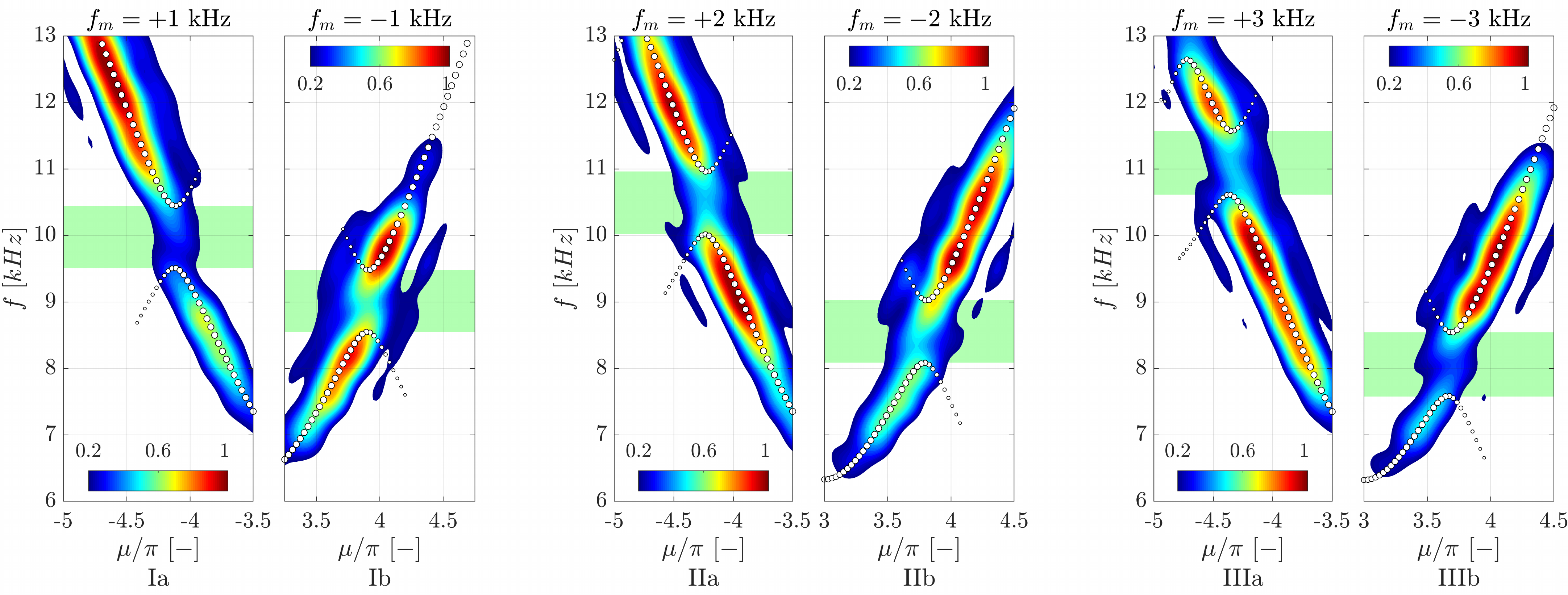} 
	\caption{Experimental (colored contours) and PWEM dispersion relations (white dots), for three levels of positive and negative switching frequencies. Experimental dispersion amplitudes are normalized by their respective maxima. On the background, directional bands as predicted by the PWEM are highlighted in light green.}	
	\label{fig:4}
\end{figure*}
The latter allow to quantify effective dimensionless modulation parameter $\alpha_m=E_m/E_{s,0}=27.5\%$, with $E_{s,0}=\left(E_{s,ON}+E_{s,OFF}\right)/2$ and $E_m=\left(E_{s,ON}-E_{s,OFF}\right)/2$. 
Traveling stiffness is achieved by phasing three spatially consecutive temporal modulation profiles, establishing material periodicity both in space and time. The ST unit cell effective stiffness, comprehending active and passive sub-lattice elements is illustrated in Fig. \ref{fig:2b} within the domain $D=\left[0,\lambda_m\right]\times\left[0,T_m\right]$, being $\lambda_m$ the spatial period. Forward (Fig. \ref{fig:2b}$-{\MakeUppercase{\romannumeral 1 }}$) and backward (Fig. \ref{fig:2b}$-{\MakeUppercase{\romannumeral 2 }}$) traveling modulations are achieved for $f_m>0$ and $f_m<0$, respectively.
Notice that, for convenience, in this work the excitation is fixed in space, therefore non-reciprocity is tested by simply changing the modulation speed from positive to negative. This is \textit{de facto} equivalent to keeping speed's sign fixed and moving the excitation source from one end to the other.\\
%
%
\begin{figure}[b]
	\centering
	\includegraphics[width=.47\textwidth]{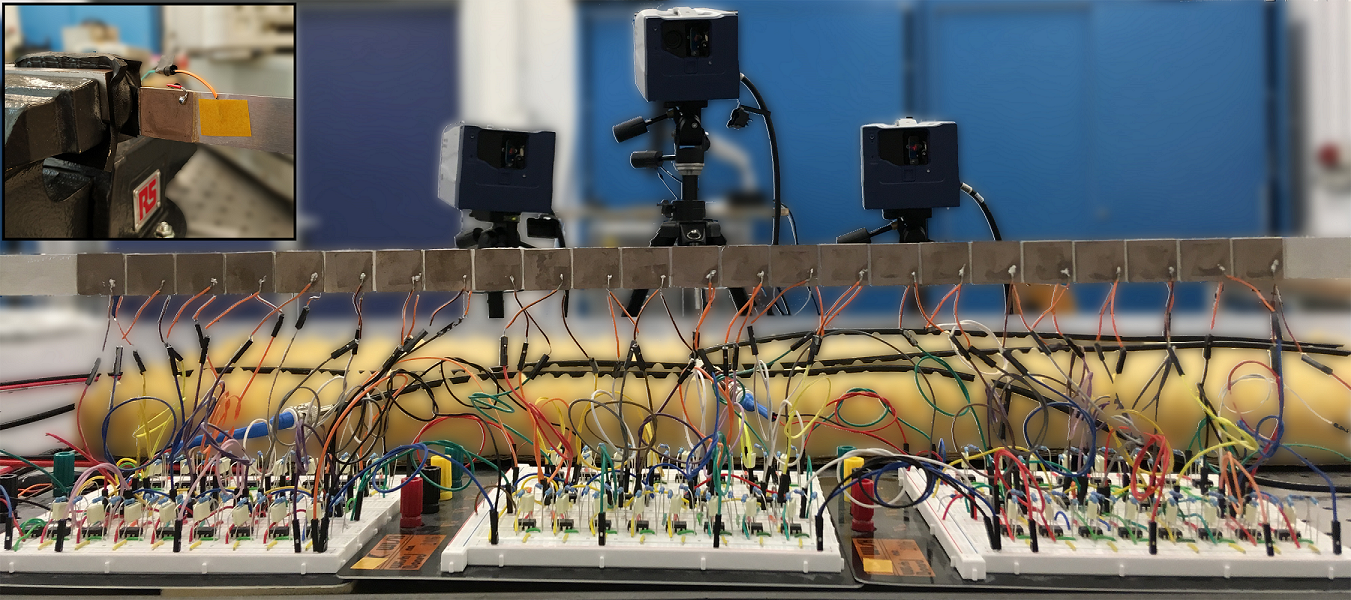}
	\caption{Front view of the piezoelectric active domain with attached negative capacitance shunts. A \textit{3D Scanner Laser Doppler Vibrometer (SLDV)} measures the velocity field along the beam.}
	\label{fig:3}
\end{figure}
\indent
For each testing condition, the experimental dispersion relation is compared with the theoretical Bloch diagrams computed through the \textit{Plane Wave Expansion Method} (PWEM). Timoshenko-beam model \cite{graff2012wave} is employed based on the governing equation for a beam:
\begin{equation}
\begin{split}
\left(EI\alpha_x\right)_x+cAG\left(w_x-\alpha\right)=\left(\rho I\alpha_{t}\right)_t\\
\left(cAG\left(w_x-\alpha\right)\right)_x=\left(\rho Aw_{t}\right)_t
\end{split}
\label{eq:01}
\end{equation}
where $\left(\cdot\right)_x=\partial\left(\cdot\right)/\partial x$ and $\left(\cdot\right)_t=\partial\left(\cdot\right)/\partial t$, $w$ is the transverse displacement of the mid-surface, $\alpha$ is the cross sectional rotation of the beam, $c=5/6$ is the shear correction coefficient, and $G$ is the shear modulus.
Given that, $EI\left(x,t\right)$, $cAG\left(x,t\right)$, $\rho I\left(x,t\right)$ and $\rho A\left(x,t\right)$ are periodic functions in space and time and can be expressed as Fourier series $C^{\left(k\right)}=\sum_{h,n=-\infty}^{+\infty}c_{h,n}^{\left(k\right)}{\rm e}^{{\rm i}\left(h\kappa_mx-n\omega_mt\right)}$ respectively, with $k=1,...,4$, being $\kappa_m=2\pi/\lambda_m$ the modulation wavenumber. Notice that in the present case all inertial properties are constant in time. Ansatz solutions are sought as propagating waves along $x$, owning same periodicity of the modulation, therefore the out of plane displacement field $w\left(x,t\right)$ and angle of rotation $\alpha\left(x,t\right)$ are approximated in terms of exponential wave functions $\alpha\left(x,t\right)=\sum_{p,q=-\infty}^{+\infty}{a}_{p,q}{\rm e}^{{\rm i}\left(p\kappa_mx-q\omega_mt\right)}{\rm e}^{{\rm i}\left(\kappa x-\omega t\right)}$ and $w\left(x,t\right)=\sum_{p,q=-\infty}^{+\infty}{b}_{p,q}{\rm e}^{{\rm i}\left(p\kappa_mx-q\omega_mt\right)}{\rm e}^{{\rm i}\left(\kappa x-\omega t\right)}$. Dispersion relations are then computed from a quadratic eigenvalue problem $\omega=\omega\left(\kappa\right)$ whose formulation is detailed in the supplementary material SM.\\
%
\begin{figure*}[t]
	\centering
	\subfloat[]{\includegraphics[width=0.42\textwidth]{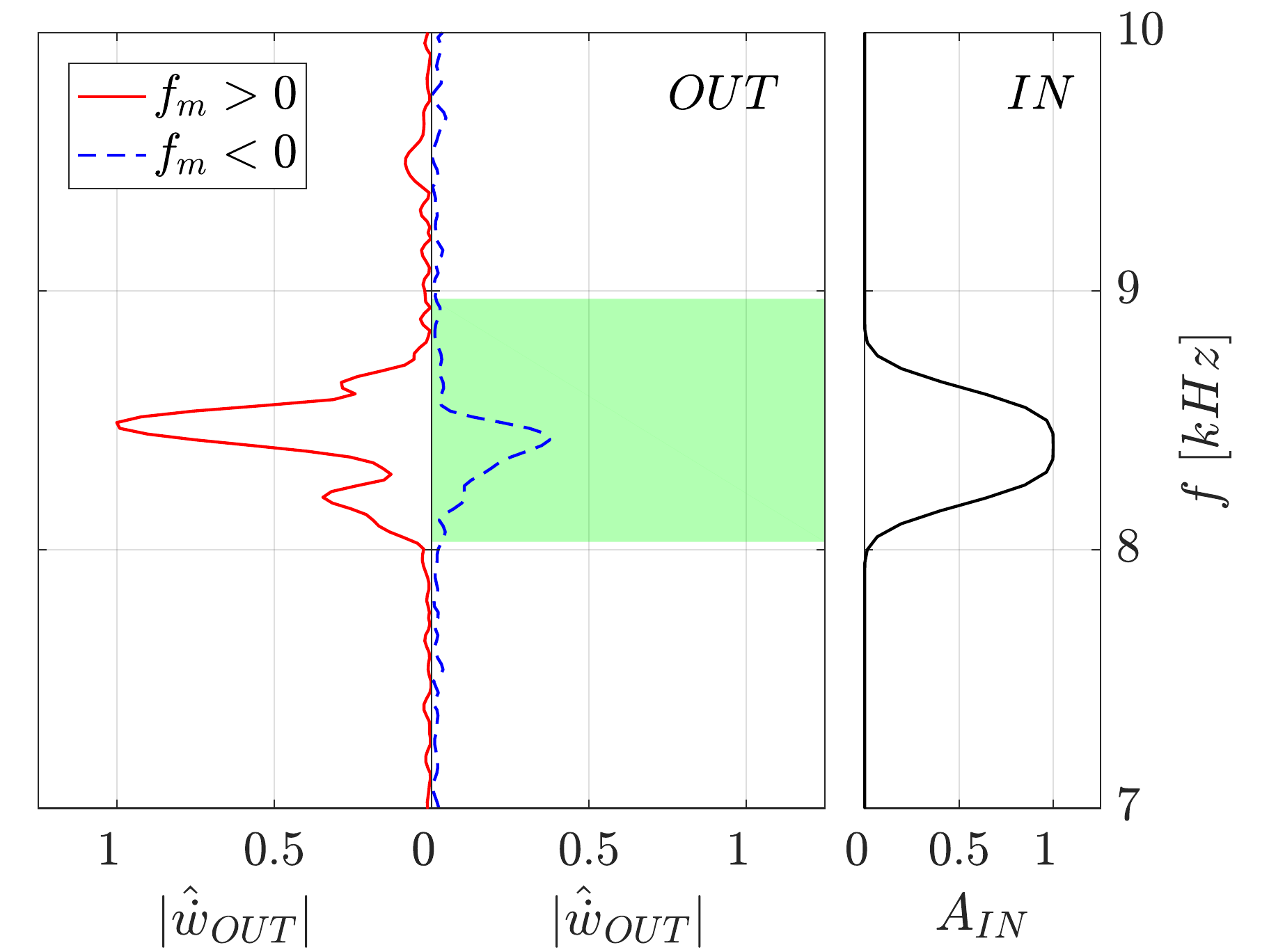} \label{fig:5a}}\hspace{0.5cm}
	\subfloat[]{\includegraphics[width=0.42\textwidth]{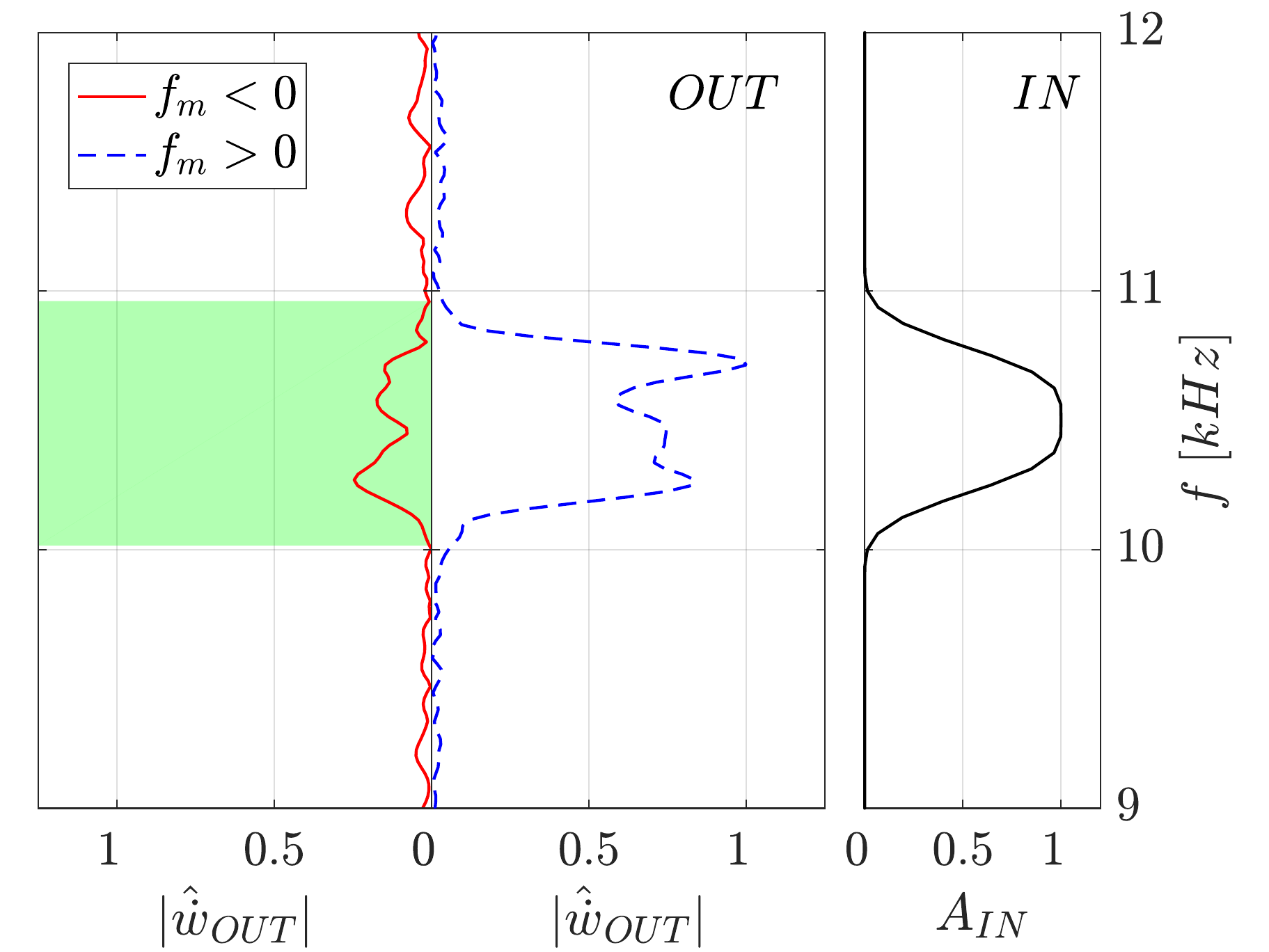} \label{fig:5b}}
	\vspace{-10pt}
	\caption{Frequency spectrum $|\hat{\dot w}_{OUT}\left(\omega\right)|$ for waves exiting the patch array for $f_m=\pm2\;{\rm kHz}$. Data is normalized by the maximum amplitude. (a) Narrowband excitation centered at $f_E=8.5\;{\rm kHz}$ with $\Delta f_{E}=0.8\;{\rm kHz}$ and corresponding input signal. (b) Excitation central frequency $f_{E}=10.5\;{\rm kHz}$ with $\Delta f_{E}=1\;{\rm kHz}$.}
	\label{fig:03}
\end{figure*}
\indent
Figure \ref{fig:3} shows the experimental setup. Input waves are excited on the left side of the beam, in correspondence of the excitation piezo pair. 
An external trigger synchronously starts the acquisition system, the external excitation and the switches of the circuits, so that each individual test is assured to be performed under the same conditions. The switches are controlled with a NI Compact-RIO and the acquisition is performed by a Polytec 3D \textit{Scanner Laser Doppler Vibrometer} (SLDV) which measures the out of plane velocity field along the beam length as the input wave propagates through the active domain.\\
\indent
Nonreciprocity is probed under two conditions. First, a wide spectrum tone burst excitation centered at bandgap frequencies is applied for analyzing positive and negative traveling modulations. Corresponding experimental dispersion relations are estimated by computing the 2D Fourier Transform of the measured velocity field $\hat{\dot w}\left(\kappa,\omega\right)=\int_{-\infty}^{+\infty}\int_{-\infty}^{+\infty}\dot{w}\left(x,t\right){\rm e}^{{\rm - i}\left(\kappa x -\omega t\right)}{\rm dxdt}$ within the domain of the structure.
When $f_m=0$, the only-space periodic medium supports a reciprocal bandgap (which is not present when all the patches are switched OFF) with a central frequency of $f_{BG}=9.5\;{\rm kHz}$ and amplitude $\Delta f\approx1\;{\rm kHz}$. When the spatiotemporal modulation is turned ON ($f_m>0$), this bandgap moves to higher frequencies (Fig. \ref{fig:4} Ia-IIa-IIIa). For $f_m<0$, opposite behavior is observed and the directional bandgap moves towards lower frequencies (Fig. \ref{fig:4} Ib-IIb-IIIb).  
We also observe that experimental data is in good agreement with the predicted analytic dispersion relation $\omega\left(\kappa\right)$, which is represented using white dots. Small discrepancies are attributed to mechanical and electrical variability of associated parameters. We can thus conclude that the system operates as expected. \\
\indent
Finally, unidirectional attenuation levels are tested using a narrow-band excitation spectrum centered at $f_E=8.5\;{\rm kHz}$ with $\Delta f_{E}=0.8\;{\rm kHz}$, a modulation frequency $f_m=\pm 2\;{\rm kHz}$ and measuring the response at $x=l_1+l$, that is just after the piezo array. This way it is possible to asses spectral content $\hat{\dot w}_{OUT}\left(\omega\right)$ of the wave packet leaving the modulated domain. Specifically, it can be noticed that for $f_m<0$ wave propagation occurs with attenuation, as shown by the blue dashed curve in Fig. \ref{fig:5a}. In contrast, a positive $f_m$ high amplitude propagating waves (red continuous curve). Analogous but reversed results are obtained when the same tests are repeated with an input wave-packet centered at $f_{E}=10.5\;{\rm kHz}$ with $\Delta f_{E}=1\;{\rm kHz}$. Wave decay is observed for positive switching frequencies (red continuous curve), while for negative values propagation is allowed (blue dashed curve).\\
\\
\indent
In conclusion, this work experimentally investigates a modulated beam with space-time periodic properties and its ensuing non-reciprocal behavior. The modulation is produced through an array of piezo patches bonded to the beam, connected to switching negative shunting circuits. Breaking time invariance, directional band-gaps and nonreciprocal  behavior for propagating waves is observed, in a frequency range spanning from $8\;{\rm kHz}$ to $11\;{\rm kHz}$.\\
Compared with other existing solutions, the proposed setup may be more easily integrated into micro electro-mechanical systems (MEMS), opening new possibilities for wave control in phononic communication devices. Moreover, we can conclude that direct piezoelectric modulation is a viable platform for the investigation of a variety physical phenomena associated with time-varying media.\\
\\
Support by the Italian Ministry of Education, University and Research, through the project Dipartimento di Eccellenza LIS4.0 (Integrated Laboratory for Lightweight and Smart Structures), is acknowledged.

\bibliography{References}

\begin{thebibliography}{32}%
\makeatletter
\providecommand \@ifxundefined [1]{%
 \@ifx{#1\undefined}
}%
\providecommand \@ifnum [1]{%
 \ifnum #1\expandafter \@firstoftwo
 \else \expandafter \@secondoftwo
 \fi
}%
\providecommand \@ifx [1]{%
 \ifx #1\expandafter \@firstoftwo
 \else \expandafter \@secondoftwo
 \fi
}%
\providecommand \natexlab [1]{#1}%
\providecommand \enquote  [1]{``#1''}%
\providecommand \bibnamefont  [1]{#1}%
\providecommand \bibfnamefont [1]{#1}%
\providecommand \citenamefont [1]{#1}%
\providecommand \href@noop [0]{\@secondoftwo}%
\providecommand \href [0]{\begingroup \@sanitize@url \@href}%
\providecommand \@href[1]{\@@startlink{#1}\@@href}%
\providecommand \@@href[1]{\endgroup#1\@@endlink}%
\providecommand \@sanitize@url [0]{\catcode `\\12\catcode `\$12\catcode
  `\&12\catcode `\#12\catcode `\^12\catcode `\_12\catcode `\%12\relax}%
\providecommand \@@startlink[1]{}%
\providecommand \@@endlink[0]{}%
\providecommand \url  [0]{\begingroup\@sanitize@url \@url }%
\providecommand \@url [1]{\endgroup\@href {#1}{\urlprefix }}%
\providecommand \urlprefix  [0]{URL }%
\providecommand \Eprint [0]{\href }%
\providecommand \doibase [0]{http://dx.doi.org/}%
\providecommand \selectlanguage [0]{\@gobble}%
\providecommand \bibinfo  [0]{\@secondoftwo}%
\providecommand \bibfield  [0]{\@secondoftwo}%
\providecommand \translation [1]{[#1]}%
\providecommand \BibitemOpen [0]{}%
\providecommand \bibitemStop [0]{}%
\providecommand \bibitemNoStop [0]{.\EOS\space}%
\providecommand \EOS [0]{\spacefactor3000\relax}%
\providecommand \BibitemShut  [1]{\csname bibitem#1\endcsname}%
\let\auto@bib@innerbib\@empty
\bibitem [{\citenamefont {Hasan}\ and\ \citenamefont
  {Kane}(2010)}]{hasan2010colloquium}%
  \BibitemOpen
  \bibfield  {author} {\bibinfo {author} {\bibfnamefont {M.~Z.}\ \bibnamefont
  {Hasan}}\ and\ \bibinfo {author} {\bibfnamefont {C.~L.}\ \bibnamefont
  {Kane}},\ }\href@noop {} {\bibfield  {journal} {\bibinfo  {journal} {Reviews
  of modern physics}\ }\textbf {\bibinfo {volume} {82}},\ \bibinfo {pages}
  {3045} (\bibinfo {year} {2010})}\BibitemShut {NoStop}%
\bibitem [{\citenamefont {Khanikaev}\ \emph {et~al.}(2013)\citenamefont
  {Khanikaev}, \citenamefont {Mousavi}, \citenamefont {Tse}, \citenamefont
  {Kargarian}, \citenamefont {MacDonald},\ and\ \citenamefont
  {Shvets}}]{khanikaev2013photonic}%
  \BibitemOpen
  \bibfield  {author} {\bibinfo {author} {\bibfnamefont {A.~B.}\ \bibnamefont
  {Khanikaev}}, \bibinfo {author} {\bibfnamefont {S.~H.}\ \bibnamefont
  {Mousavi}}, \bibinfo {author} {\bibfnamefont {W.-K.}\ \bibnamefont {Tse}},
  \bibinfo {author} {\bibfnamefont {M.}~\bibnamefont {Kargarian}}, \bibinfo
  {author} {\bibfnamefont {A.~H.}\ \bibnamefont {MacDonald}}, \ and\ \bibinfo
  {author} {\bibfnamefont {G.}~\bibnamefont {Shvets}},\ }\href@noop {}
  {\bibfield  {journal} {\bibinfo  {journal} {Nature materials}\ }\textbf
  {\bibinfo {volume} {12}},\ \bibinfo {pages} {233} (\bibinfo {year}
  {2013})}\BibitemShut {NoStop}%
\bibitem [{\citenamefont {Chamanara}\ \emph {et~al.}(2017)\citenamefont
  {Chamanara}, \citenamefont {Taravati}, \citenamefont {Deck-L{\'e}ger},\ and\
  \citenamefont {Caloz}}]{chamanara2017optical}%
  \BibitemOpen
  \bibfield  {author} {\bibinfo {author} {\bibfnamefont {N.}~\bibnamefont
  {Chamanara}}, \bibinfo {author} {\bibfnamefont {S.}~\bibnamefont {Taravati}},
  \bibinfo {author} {\bibfnamefont {Z.-L.}\ \bibnamefont {Deck-L{\'e}ger}}, \
  and\ \bibinfo {author} {\bibfnamefont {C.}~\bibnamefont {Caloz}},\
  }\href@noop {} {\bibfield  {journal} {\bibinfo  {journal} {Physical Review
  B}\ }\textbf {\bibinfo {volume} {96}},\ \bibinfo {pages} {155409} (\bibinfo
  {year} {2017})}\BibitemShut {NoStop}%
\bibitem [{\citenamefont {Cummer}\ \emph {et~al.}(2016)\citenamefont {Cummer},
  \citenamefont {Christensen},\ and\ \citenamefont
  {Al{\`u}}}]{cummer2016controlling}%
  \BibitemOpen
  \bibfield  {author} {\bibinfo {author} {\bibfnamefont {S.~A.}\ \bibnamefont
  {Cummer}}, \bibinfo {author} {\bibfnamefont {J.}~\bibnamefont {Christensen}},
  \ and\ \bibinfo {author} {\bibfnamefont {A.}~\bibnamefont {Al{\`u}}},\
  }\href@noop {} {\bibfield  {journal} {\bibinfo  {journal} {Nature Reviews
  Materials}\ }\textbf {\bibinfo {volume} {1}},\ \bibinfo {pages} {16001}
  (\bibinfo {year} {2016})}\BibitemShut {NoStop}%
\bibitem [{\citenamefont {Li}\ \emph {et~al.}(2019)\citenamefont {Li},
  \citenamefont {Shen}, \citenamefont {Zhu}, \citenamefont {Xie},\ and\
  \citenamefont {Cummer}}]{li2019nonreciprocal}%
  \BibitemOpen
  \bibfield  {author} {\bibinfo {author} {\bibfnamefont {J.}~\bibnamefont
  {Li}}, \bibinfo {author} {\bibfnamefont {C.}~\bibnamefont {Shen}}, \bibinfo
  {author} {\bibfnamefont {X.}~\bibnamefont {Zhu}}, \bibinfo {author}
  {\bibfnamefont {Y.}~\bibnamefont {Xie}}, \ and\ \bibinfo {author}
  {\bibfnamefont {S.~A.}\ \bibnamefont {Cummer}},\ }\href@noop {} {\bibfield
  {journal} {\bibinfo  {journal} {Physical Review B}\ }\textbf {\bibinfo
  {volume} {99}},\ \bibinfo {pages} {144311} (\bibinfo {year}
  {2019})}\BibitemShut {NoStop}%
\bibitem [{\citenamefont {Karkar}\ \emph {et~al.}(2019)\citenamefont {Karkar},
  \citenamefont {De~Bono}, \citenamefont {Collet}, \citenamefont {Matten},
  \citenamefont {Ouisse},\ and\ \citenamefont {Rivet}}]{karkar2019broadband}%
  \BibitemOpen
  \bibfield  {author} {\bibinfo {author} {\bibfnamefont {S.}~\bibnamefont
  {Karkar}}, \bibinfo {author} {\bibfnamefont {E.}~\bibnamefont {De~Bono}},
  \bibinfo {author} {\bibfnamefont {M.}~\bibnamefont {Collet}}, \bibinfo
  {author} {\bibfnamefont {G.}~\bibnamefont {Matten}}, \bibinfo {author}
  {\bibfnamefont {M.}~\bibnamefont {Ouisse}}, \ and\ \bibinfo {author}
  {\bibfnamefont {E.}~\bibnamefont {Rivet}},\ }\href@noop {} {\bibfield
  {journal} {\bibinfo  {journal} {arXiv preprint arXiv:1906.09099}\ } (\bibinfo
  {year} {2019})}\BibitemShut {NoStop}%
\bibitem [{\citenamefont {Vila}\ \emph {et~al.}(2017)\citenamefont {Vila},
  \citenamefont {Pal}, \citenamefont {Ruzzene},\ and\ \citenamefont
  {Trainiti}}]{vila2017bloch}%
  \BibitemOpen
  \bibfield  {author} {\bibinfo {author} {\bibfnamefont {J.}~\bibnamefont
  {Vila}}, \bibinfo {author} {\bibfnamefont {R.~K.}\ \bibnamefont {Pal}},
  \bibinfo {author} {\bibfnamefont {M.}~\bibnamefont {Ruzzene}}, \ and\
  \bibinfo {author} {\bibfnamefont {G.}~\bibnamefont {Trainiti}},\ }\href@noop
  {} {\bibfield  {journal} {\bibinfo  {journal} {Journal of Sound and
  Vibration}\ }\textbf {\bibinfo {volume} {406}},\ \bibinfo {pages} {363}
  (\bibinfo {year} {2017})}\BibitemShut {NoStop}%
\bibitem [{\citenamefont {Nassar}\ \emph
  {et~al.}(2017{\natexlab{a}})\citenamefont {Nassar}, \citenamefont {Xu},
  \citenamefont {Norris},\ and\ \citenamefont {Huang}}]{nassar2017modulated}%
  \BibitemOpen
  \bibfield  {author} {\bibinfo {author} {\bibfnamefont {H.}~\bibnamefont
  {Nassar}}, \bibinfo {author} {\bibfnamefont {X.}~\bibnamefont {Xu}}, \bibinfo
  {author} {\bibfnamefont {A.}~\bibnamefont {Norris}}, \ and\ \bibinfo {author}
  {\bibfnamefont {G.}~\bibnamefont {Huang}},\ }\href@noop {} {\bibfield
  {journal} {\bibinfo  {journal} {Journal of the Mechanics and Physics of
  Solids}\ }\textbf {\bibinfo {volume} {101}},\ \bibinfo {pages} {10} (\bibinfo
  {year} {2017}{\natexlab{a}})}\BibitemShut {NoStop}%
\bibitem [{\citenamefont {Wallen}\ and\ \citenamefont
  {Haberman}(2019)}]{wallen2019nonreciprocal}%
  \BibitemOpen
  \bibfield  {author} {\bibinfo {author} {\bibfnamefont {S.~P.}\ \bibnamefont
  {Wallen}}\ and\ \bibinfo {author} {\bibfnamefont {M.~R.}\ \bibnamefont
  {Haberman}},\ }\href@noop {} {\bibfield  {journal} {\bibinfo  {journal}
  {Physical Review E}\ }\textbf {\bibinfo {volume} {99}},\ \bibinfo {pages}
  {013001} (\bibinfo {year} {2019})}\BibitemShut {NoStop}%
\bibitem [{\citenamefont {Huber}(2016)}]{huber2016topological}%
  \BibitemOpen
  \bibfield  {author} {\bibinfo {author} {\bibfnamefont {S.~D.}\ \bibnamefont
  {Huber}},\ }\href@noop {} {\bibfield  {journal} {\bibinfo  {journal} {Nature
  Physics}\ }\textbf {\bibinfo {volume} {12}},\ \bibinfo {pages} {621}
  (\bibinfo {year} {2016})}\BibitemShut {NoStop}%
\bibitem [{\citenamefont {Trainiti}\ and\ \citenamefont
  {Ruzzene}(2016)}]{trainiti2016non}%
  \BibitemOpen
  \bibfield  {author} {\bibinfo {author} {\bibfnamefont {G.}~\bibnamefont
  {Trainiti}}\ and\ \bibinfo {author} {\bibfnamefont {M.}~\bibnamefont
  {Ruzzene}},\ }\href@noop {} {\bibfield  {journal} {\bibinfo  {journal} {New
  Journal of Physics}\ }\textbf {\bibinfo {volume} {18}},\ \bibinfo {pages}
  {083047} (\bibinfo {year} {2016})}\BibitemShut {NoStop}%
\bibitem [{\citenamefont {Yi}\ \emph {et~al.}(2018)\citenamefont {Yi},
  \citenamefont {Collet},\ and\ \citenamefont {Karkar}}]{yi2018reflection}%
  \BibitemOpen
  \bibfield  {author} {\bibinfo {author} {\bibfnamefont {K.}~\bibnamefont
  {Yi}}, \bibinfo {author} {\bibfnamefont {M.}~\bibnamefont {Collet}}, \ and\
  \bibinfo {author} {\bibfnamefont {S.}~\bibnamefont {Karkar}},\ }\href@noop {}
  {\bibfield  {journal} {\bibinfo  {journal} {Physical Review B}\ }\textbf
  {\bibinfo {volume} {98}},\ \bibinfo {pages} {054109} (\bibinfo {year}
  {2018})}\BibitemShut {NoStop}%
\bibitem [{\citenamefont {Oudich}\ \emph {et~al.}(2019)\citenamefont {Oudich},
  \citenamefont {Deng}, \citenamefont {Tao},\ and\ \citenamefont
  {Jing}}]{oudich2019space}%
  \BibitemOpen
  \bibfield  {author} {\bibinfo {author} {\bibfnamefont {M.}~\bibnamefont
  {Oudich}}, \bibinfo {author} {\bibfnamefont {Y.}~\bibnamefont {Deng}},
  \bibinfo {author} {\bibfnamefont {M.}~\bibnamefont {Tao}}, \ and\ \bibinfo
  {author} {\bibfnamefont {Y.}~\bibnamefont {Jing}},\ }\href@noop {} {\bibfield
   {journal} {\bibinfo  {journal} {arXiv preprint arXiv:1904.02711}\ }
  (\bibinfo {year} {2019})}\BibitemShut {NoStop}%
\bibitem [{\citenamefont {Fleury}\ \emph {et~al.}(2014)\citenamefont {Fleury},
  \citenamefont {Sounas}, \citenamefont {Sieck}, \citenamefont {Haberman},\
  and\ \citenamefont {Al{\`u}}}]{fleury2014sound}%
  \BibitemOpen
  \bibfield  {author} {\bibinfo {author} {\bibfnamefont {R.}~\bibnamefont
  {Fleury}}, \bibinfo {author} {\bibfnamefont {D.~L.}\ \bibnamefont {Sounas}},
  \bibinfo {author} {\bibfnamefont {C.~F.}\ \bibnamefont {Sieck}}, \bibinfo
  {author} {\bibfnamefont {M.~R.}\ \bibnamefont {Haberman}}, \ and\ \bibinfo
  {author} {\bibfnamefont {A.}~\bibnamefont {Al{\`u}}},\ }\href@noop {}
  {\bibfield  {journal} {\bibinfo  {journal} {Science}\ }\textbf {\bibinfo
  {volume} {343}},\ \bibinfo {pages} {516} (\bibinfo {year}
  {2014})}\BibitemShut {NoStop}%
\bibitem [{\citenamefont {Fleury}\ \emph {et~al.}(2015)\citenamefont {Fleury},
  \citenamefont {Sounas},\ and\ \citenamefont
  {Al{\`u}}}]{fleury2015subwavelength}%
  \BibitemOpen
  \bibfield  {author} {\bibinfo {author} {\bibfnamefont {R.}~\bibnamefont
  {Fleury}}, \bibinfo {author} {\bibfnamefont {D.~L.}\ \bibnamefont {Sounas}},
  \ and\ \bibinfo {author} {\bibfnamefont {A.}~\bibnamefont {Al{\`u}}},\
  }\href@noop {} {\bibfield  {journal} {\bibinfo  {journal} {Physical Review
  B}\ }\textbf {\bibinfo {volume} {91}},\ \bibinfo {pages} {174306} (\bibinfo
  {year} {2015})}\BibitemShut {NoStop}%
\bibitem [{\citenamefont {Khanikaev}\ \emph {et~al.}(2015)\citenamefont
  {Khanikaev}, \citenamefont {Fleury}, \citenamefont {Mousavi},\ and\
  \citenamefont {Alu}}]{khanikaev2015topologically}%
  \BibitemOpen
  \bibfield  {author} {\bibinfo {author} {\bibfnamefont {A.~B.}\ \bibnamefont
  {Khanikaev}}, \bibinfo {author} {\bibfnamefont {R.}~\bibnamefont {Fleury}},
  \bibinfo {author} {\bibfnamefont {S.~H.}\ \bibnamefont {Mousavi}}, \ and\
  \bibinfo {author} {\bibfnamefont {A.}~\bibnamefont {Alu}},\ }\href@noop {}
  {\bibfield  {journal} {\bibinfo  {journal} {Nature communications}\ }\textbf
  {\bibinfo {volume} {6}},\ \bibinfo {pages} {8260} (\bibinfo {year}
  {2015})}\BibitemShut {NoStop}%
\bibitem [{\citenamefont {Chen}\ \emph
  {et~al.}(2019{\natexlab{a}})\citenamefont {Chen}, \citenamefont {Yao},
  \citenamefont {Nassar},\ and\ \citenamefont {Huang}}]{chen2019mechanical}%
  \BibitemOpen
  \bibfield  {author} {\bibinfo {author} {\bibfnamefont {H.}~\bibnamefont
  {Chen}}, \bibinfo {author} {\bibfnamefont {L.}~\bibnamefont {Yao}}, \bibinfo
  {author} {\bibfnamefont {H.}~\bibnamefont {Nassar}}, \ and\ \bibinfo {author}
  {\bibfnamefont {G.}~\bibnamefont {Huang}},\ }\href@noop {} {\bibfield
  {journal} {\bibinfo  {journal} {Physical Review Applied}\ }\textbf {\bibinfo
  {volume} {11}},\ \bibinfo {pages} {044029} (\bibinfo {year}
  {2019}{\natexlab{a}})}\BibitemShut {NoStop}%
\bibitem [{\citenamefont {Wang}\ \emph {et~al.}(2015)\citenamefont {Wang},
  \citenamefont {Lu},\ and\ \citenamefont {Bertoldi}}]{wang2015topological}%
  \BibitemOpen
  \bibfield  {author} {\bibinfo {author} {\bibfnamefont {P.}~\bibnamefont
  {Wang}}, \bibinfo {author} {\bibfnamefont {L.}~\bibnamefont {Lu}}, \ and\
  \bibinfo {author} {\bibfnamefont {K.}~\bibnamefont {Bertoldi}},\ }\href@noop
  {} {\bibfield  {journal} {\bibinfo  {journal} {Physical review letters}\
  }\textbf {\bibinfo {volume} {115}},\ \bibinfo {pages} {104302} (\bibinfo
  {year} {2015})}\BibitemShut {NoStop}%
\bibitem [{\citenamefont {Ni}\ \emph {et~al.}(2015)\citenamefont {Ni},
  \citenamefont {He}, \citenamefont {Sun}, \citenamefont {Liu}, \citenamefont
  {Lu}, \citenamefont {Feng},\ and\ \citenamefont
  {Chen}}]{ni2015topologically}%
  \BibitemOpen
  \bibfield  {author} {\bibinfo {author} {\bibfnamefont {X.}~\bibnamefont
  {Ni}}, \bibinfo {author} {\bibfnamefont {C.}~\bibnamefont {He}}, \bibinfo
  {author} {\bibfnamefont {X.-C.}\ \bibnamefont {Sun}}, \bibinfo {author}
  {\bibfnamefont {X.-p.}\ \bibnamefont {Liu}}, \bibinfo {author} {\bibfnamefont
  {M.-H.}\ \bibnamefont {Lu}}, \bibinfo {author} {\bibfnamefont
  {L.}~\bibnamefont {Feng}}, \ and\ \bibinfo {author} {\bibfnamefont {Y.-F.}\
  \bibnamefont {Chen}},\ }\href@noop {} {\bibfield  {journal} {\bibinfo
  {journal} {New Journal of Physics}\ }\textbf {\bibinfo {volume} {17}},\
  \bibinfo {pages} {053016} (\bibinfo {year} {2015})}\BibitemShut {NoStop}%
\bibitem [{\citenamefont {Nash}\ \emph {et~al.}(2015)\citenamefont {Nash},
  \citenamefont {Kleckner}, \citenamefont {Read}, \citenamefont {Vitelli},
  \citenamefont {Turner},\ and\ \citenamefont {Irvine}}]{nash2015topological}%
  \BibitemOpen
  \bibfield  {author} {\bibinfo {author} {\bibfnamefont {L.~M.}\ \bibnamefont
  {Nash}}, \bibinfo {author} {\bibfnamefont {D.}~\bibnamefont {Kleckner}},
  \bibinfo {author} {\bibfnamefont {A.}~\bibnamefont {Read}}, \bibinfo {author}
  {\bibfnamefont {V.}~\bibnamefont {Vitelli}}, \bibinfo {author} {\bibfnamefont
  {A.~M.}\ \bibnamefont {Turner}}, \ and\ \bibinfo {author} {\bibfnamefont
  {W.~T.}\ \bibnamefont {Irvine}},\ }\href@noop {} {\bibfield  {journal}
  {\bibinfo  {journal} {Proceedings of the National Academy of Sciences}\
  }\textbf {\bibinfo {volume} {112}},\ \bibinfo {pages} {14495} (\bibinfo
  {year} {2015})}\BibitemShut {NoStop}%
\bibitem [{\citenamefont {Mojahed}\ \emph {et~al.}(2019)\citenamefont
  {Mojahed}, \citenamefont {Gendelman},\ and\ \citenamefont
  {Vakakis}}]{Mojahed2019}%
  \BibitemOpen
  \bibfield  {author} {\bibinfo {author} {\bibfnamefont {A.}~\bibnamefont
  {Mojahed}}, \bibinfo {author} {\bibfnamefont {O.~V.}\ \bibnamefont
  {Gendelman}}, \ and\ \bibinfo {author} {\bibfnamefont {A.~F.}\ \bibnamefont
  {Vakakis}},\ }\href {\doibase 10.1121/1.5114915} {\bibfield  {journal}
  {\bibinfo  {journal} {The Journal of the Acoustical Society of America}\
  }\textbf {\bibinfo {volume} {146}},\ \bibinfo {pages} {826} (\bibinfo {year}
  {2019})}\BibitemShut {NoStop}%
\bibitem [{\citenamefont {Fronk}\ \emph {et~al.}(2019)\citenamefont {Fronk},
  \citenamefont {Tawfick}, \citenamefont {Daraio}, \citenamefont {Li},
  \citenamefont {Vakakis},\ and\ \citenamefont {Leamy}}]{Fronk2019}%
  \BibitemOpen
  \bibfield  {author} {\bibinfo {author} {\bibfnamefont {M.~D.}\ \bibnamefont
  {Fronk}}, \bibinfo {author} {\bibfnamefont {S.}~\bibnamefont {Tawfick}},
  \bibinfo {author} {\bibfnamefont {C.}~\bibnamefont {Daraio}}, \bibinfo
  {author} {\bibfnamefont {S.}~\bibnamefont {Li}}, \bibinfo {author}
  {\bibfnamefont {A.}~\bibnamefont {Vakakis}}, \ and\ \bibinfo {author}
  {\bibfnamefont {M.~J.}\ \bibnamefont {Leamy}},\ }\href {\doibase
  10.1115/1.4043783} {\bibfield  {journal} {\bibinfo  {journal} {Journal of
  Vibration and Acoustics, Transactions of the ASME}\ }\textbf {\bibinfo
  {volume} {141}},\ \bibinfo {pages} {1} (\bibinfo {year} {2019})}\BibitemShut
  {NoStop}%
\bibitem [{\citenamefont {Wu}\ \emph {et~al.}(2018)\citenamefont {Wu},
  \citenamefont {Zheng},\ and\ \citenamefont {Wang}}]{wu2018metastable}%
  \BibitemOpen
  \bibfield  {author} {\bibinfo {author} {\bibfnamefont {Z.}~\bibnamefont
  {Wu}}, \bibinfo {author} {\bibfnamefont {Y.}~\bibnamefont {Zheng}}, \ and\
  \bibinfo {author} {\bibfnamefont {K.}~\bibnamefont {Wang}},\ }\href@noop {}
  {\bibfield  {journal} {\bibinfo  {journal} {Physical Review E}\ }\textbf
  {\bibinfo {volume} {97}},\ \bibinfo {pages} {022209} (\bibinfo {year}
  {2018})}\BibitemShut {NoStop}%
\bibitem [{\citenamefont {Boechler}\ \emph {et~al.}(2011)\citenamefont
  {Boechler}, \citenamefont {Theocharis},\ and\ \citenamefont
  {Daraio}}]{boechler2011bifurcation}%
  \BibitemOpen
  \bibfield  {author} {\bibinfo {author} {\bibfnamefont {N.}~\bibnamefont
  {Boechler}}, \bibinfo {author} {\bibfnamefont {G.}~\bibnamefont
  {Theocharis}}, \ and\ \bibinfo {author} {\bibfnamefont {C.}~\bibnamefont
  {Daraio}},\ }\href@noop {} {\bibfield  {journal} {\bibinfo  {journal} {Nature
  materials}\ }\textbf {\bibinfo {volume} {10}},\ \bibinfo {pages} {665}
  (\bibinfo {year} {2011})}\BibitemShut {NoStop}%
\bibitem [{\citenamefont {Nassar}\ \emph
  {et~al.}(2017{\natexlab{b}})\citenamefont {Nassar}, \citenamefont {Chen},
  \citenamefont {Norris}, \citenamefont {Haberman},\ and\ \citenamefont
  {Huang}}]{nassar2017non}%
  \BibitemOpen
  \bibfield  {author} {\bibinfo {author} {\bibfnamefont {H.}~\bibnamefont
  {Nassar}}, \bibinfo {author} {\bibfnamefont {H.}~\bibnamefont {Chen}},
  \bibinfo {author} {\bibfnamefont {A.}~\bibnamefont {Norris}}, \bibinfo
  {author} {\bibfnamefont {M.}~\bibnamefont {Haberman}}, \ and\ \bibinfo
  {author} {\bibfnamefont {G.}~\bibnamefont {Huang}},\ }\href@noop {}
  {\bibfield  {journal} {\bibinfo  {journal} {Proceedings of the Royal Society
  A: Mathematical, Physical and Engineering Sciences}\ }\textbf {\bibinfo
  {volume} {473}},\ \bibinfo {pages} {20170188} (\bibinfo {year}
  {2017}{\natexlab{b}})}\BibitemShut {NoStop}%
\bibitem [{\citenamefont {Riva}\ \emph {et~al.}(2019)\citenamefont {Riva},
  \citenamefont {Marconi}, \citenamefont {Cazzulani},\ and\ \citenamefont
  {Braghin}}]{riva2019generalized}%
  \BibitemOpen
  \bibfield  {author} {\bibinfo {author} {\bibfnamefont {E.}~\bibnamefont
  {Riva}}, \bibinfo {author} {\bibfnamefont {J.}~\bibnamefont {Marconi}},
  \bibinfo {author} {\bibfnamefont {G.}~\bibnamefont {Cazzulani}}, \ and\
  \bibinfo {author} {\bibfnamefont {F.}~\bibnamefont {Braghin}},\ }\href@noop
  {} {\bibfield  {journal} {\bibinfo  {journal} {Journal of Sound and
  Vibration}\ }\textbf {\bibinfo {volume} {449}},\ \bibinfo {pages} {172}
  (\bibinfo {year} {2019})}\BibitemShut {NoStop}%
\bibitem [{\citenamefont {Chen}\ \emph
  {et~al.}(2019{\natexlab{b}})\citenamefont {Chen}, \citenamefont {Li},
  \citenamefont {Nassar}, \citenamefont {Norris}, \citenamefont {Daraio},\ and\
  \citenamefont {Huang}}]{chen2019nonreciprocal}%
  \BibitemOpen
  \bibfield  {author} {\bibinfo {author} {\bibfnamefont {Y.}~\bibnamefont
  {Chen}}, \bibinfo {author} {\bibfnamefont {X.}~\bibnamefont {Li}}, \bibinfo
  {author} {\bibfnamefont {H.}~\bibnamefont {Nassar}}, \bibinfo {author}
  {\bibfnamefont {A.~N.}\ \bibnamefont {Norris}}, \bibinfo {author}
  {\bibfnamefont {C.}~\bibnamefont {Daraio}}, \ and\ \bibinfo {author}
  {\bibfnamefont {G.}~\bibnamefont {Huang}},\ }\href@noop {} {\bibfield
  {journal} {\bibinfo  {journal} {Physical Review Applied}\ }\textbf {\bibinfo
  {volume} {11}},\ \bibinfo {pages} {064052} (\bibinfo {year}
  {2019}{\natexlab{b}})}\BibitemShut {NoStop}%
\bibitem [{\citenamefont {Wang}\ \emph {et~al.}(2018)\citenamefont {Wang},
  \citenamefont {Yousefzadeh}, \citenamefont {Chen}, \citenamefont {Nassar},
  \citenamefont {Huang},\ and\ \citenamefont {Daraio}}]{wang2018observation}%
  \BibitemOpen
  \bibfield  {author} {\bibinfo {author} {\bibfnamefont {Y.}~\bibnamefont
  {Wang}}, \bibinfo {author} {\bibfnamefont {B.}~\bibnamefont {Yousefzadeh}},
  \bibinfo {author} {\bibfnamefont {H.}~\bibnamefont {Chen}}, \bibinfo {author}
  {\bibfnamefont {H.}~\bibnamefont {Nassar}}, \bibinfo {author} {\bibfnamefont
  {G.}~\bibnamefont {Huang}}, \ and\ \bibinfo {author} {\bibfnamefont
  {C.}~\bibnamefont {Daraio}},\ }\href@noop {} {\bibfield  {journal} {\bibinfo
  {journal} {Physical review letters}\ }\textbf {\bibinfo {volume} {121}},\
  \bibinfo {pages} {194301} (\bibinfo {year} {2018})}\BibitemShut {NoStop}%
\bibitem [{\citenamefont {Trainiti}\ \emph {et~al.}(2019)\citenamefont
  {Trainiti}, \citenamefont {Xia}, \citenamefont {Marconi}, \citenamefont
  {Cazzulani}, \citenamefont {Erturk},\ and\ \citenamefont
  {Ruzzene}}]{trainiti2019time}%
  \BibitemOpen
  \bibfield  {author} {\bibinfo {author} {\bibfnamefont {G.}~\bibnamefont
  {Trainiti}}, \bibinfo {author} {\bibfnamefont {Y.}~\bibnamefont {Xia}},
  \bibinfo {author} {\bibfnamefont {J.}~\bibnamefont {Marconi}}, \bibinfo
  {author} {\bibfnamefont {G.}~\bibnamefont {Cazzulani}}, \bibinfo {author}
  {\bibfnamefont {A.}~\bibnamefont {Erturk}}, \ and\ \bibinfo {author}
  {\bibfnamefont {M.}~\bibnamefont {Ruzzene}},\ }\href@noop {} {\bibfield
  {journal} {\bibinfo  {journal} {Physical review letters}\ }\textbf {\bibinfo
  {volume} {122}},\ \bibinfo {pages} {124301} (\bibinfo {year}
  {2019})}\BibitemShut {NoStop}%
\bibitem [{\citenamefont {Kraus}\ \emph {et~al.}(2012)\citenamefont {Kraus},
  \citenamefont {Lahini}, \citenamefont {Ringel}, \citenamefont {Verbin},\ and\
  \citenamefont {Zilberberg}}]{kraus2012topological}%
  \BibitemOpen
  \bibfield  {author} {\bibinfo {author} {\bibfnamefont {Y.~E.}\ \bibnamefont
  {Kraus}}, \bibinfo {author} {\bibfnamefont {Y.}~\bibnamefont {Lahini}},
  \bibinfo {author} {\bibfnamefont {Z.}~\bibnamefont {Ringel}}, \bibinfo
  {author} {\bibfnamefont {M.}~\bibnamefont {Verbin}}, \ and\ \bibinfo {author}
  {\bibfnamefont {O.}~\bibnamefont {Zilberberg}},\ }\href@noop {} {\bibfield
  {journal} {\bibinfo  {journal} {Physical review letters}\ }\textbf {\bibinfo
  {volume} {109}},\ \bibinfo {pages} {106402} (\bibinfo {year}
  {2012})}\BibitemShut {NoStop}%
\bibitem [{\citenamefont {Grinberg}\ \emph {et~al.}(2019)\citenamefont
  {Grinberg}, \citenamefont {Lin}, \citenamefont {Harris}, \citenamefont
  {Benalcazar}, \citenamefont {Peterson}, \citenamefont {Hughes},\ and\
  \citenamefont {Bahl}}]{grinberg2019robust}%
  \BibitemOpen
  \bibfield  {author} {\bibinfo {author} {\bibfnamefont {I.~H.}\ \bibnamefont
  {Grinberg}}, \bibinfo {author} {\bibfnamefont {M.}~\bibnamefont {Lin}},
  \bibinfo {author} {\bibfnamefont {C.}~\bibnamefont {Harris}}, \bibinfo
  {author} {\bibfnamefont {W.~A.}\ \bibnamefont {Benalcazar}}, \bibinfo
  {author} {\bibfnamefont {C.~W.}\ \bibnamefont {Peterson}}, \bibinfo {author}
  {\bibfnamefont {T.~L.}\ \bibnamefont {Hughes}}, \ and\ \bibinfo {author}
  {\bibfnamefont {G.}~\bibnamefont {Bahl}},\ }\href@noop {} {\bibfield
  {journal} {\bibinfo  {journal} {arXiv preprint arXiv:1905.02778}\ } (\bibinfo
  {year} {2019})}\BibitemShut {NoStop}%
\bibitem [{\citenamefont {Graff}(2012)}]{graff2012wave}%
  \BibitemOpen
  \bibfield  {author} {\bibinfo {author} {\bibfnamefont {K.~F.}\ \bibnamefont
  {Graff}},\ }\href@noop {} {\emph {\bibinfo {title} {Wave motion in elastic
  solids}}}\ (\bibinfo  {publisher} {Courier Corporation},\ \bibinfo {year}
  {2012})\BibitemShut {NoStop}%
\end{thebibliography}%

\end{document}